\documentclass[journal=nalefd,manuscript=letter]{achemso}

\usepackage{graphicx}
\usepackage{dcolumn}
\usepackage{bm}
\usepackage{graphicx}
\usepackage{subfig}
\usepackage{comment}
\usepackage{amsmath}
\usepackage{amssymb}
\usepackage[below]{placeins}
\usepackage[usenames, dvipsnames]{color}
\usepackage{siunitx}
\usepackage{float}
\usepackage{tabularx}
\usepackage{hyperref}
\hypersetup{
colorlinks   = true, 
  urlcolor     = blue, 
  linkcolor    = black, 
  citecolor   = black 
}

\title{Demonstration of a Thermally-Coupled Row-Column SNSPD Imaging Array}

\author{Jason P. Allmaras}
\email{jallmara@caltech.edu}
\affiliation{Department of Applied Physics and Materials Science, California Institute of Technology, Pasadena, CA, United States.}
\alsoaffiliation{Jet Propulsion Laboratory, California Institute of Technology, Pasadena, CA, United States.}

\author{Emma E. Wollman}
\affiliation{Jet Propulsion Laboratory, California Institute of Technology, Pasadena, CA, United States.}

\author{Andrew D. Beyer}
\affiliation{Jet Propulsion Laboratory, California Institute of Technology, Pasadena, CA, United States.}

\author{Ryan M. Briggs}
\affiliation{Jet Propulsion Laboratory, California Institute of Technology, Pasadena, CA, United States.}

\author{Boris A. Korzh}
\affiliation{Jet Propulsion Laboratory, California Institute of Technology, Pasadena, CA, United States.}

\author{Bruce Bumble}
\affiliation{Jet Propulsion Laboratory, California Institute of Technology, Pasadena, CA, United States.}

\author{Matthew D. Shaw}
\affiliation{Jet Propulsion Laboratory, California Institute of Technology, Pasadena, CA, United States.}


\begin{document}
This document is the unedited Author's version of a Submitted Work that was subsequently accepted for publication in Nano Letters, copyright \copyright American Chemical Society after peer review. To access the final edited and published work see \url{http://dx.doi.org/10.1021/acs.nanolett.0c00246}.

\copyright{2020. California Institute of Technology. Government sponsorship acknowledged.}

\begin{abstract}
While single-pixel superconducting nanowire single photon detectors (SNSPDs) have demonstrated remarkable efficiency and timing performance from the UV to near-IR, scaling these devices to large imaging arrays remains challenging. Here, we propose a new SNSPD multiplexing system using thermal coupling and detection correlations between two photosensitive layers of an array. Using this architecture with the channels of one layer oriented in rows and the second layer in columns, we demonstrate imaging capability in 16-pixel arrays with accurate spot tracking at the few photon level. We also explore the performance tradeoffs of orienting the top layer nanowires parallel and perpendicular to the bottom layer.  The thermally-coupled row-column scheme is readily able to scale to the kilopixel size with existing readout systems, and when combined with other multiplexing architectures, has the potential to enable megapixel scale SNSPD imaging arrays.
\end{abstract}

Single-pixel superconducting nanowire single photon detectors (SNSPDs) \cite{goltsman_picosecond_2001} have achieved remarkable performance including \(>\)90\% detection efficiency  \cite{marsili_detecting_2013, zhang_nbn_2017}, timing jitter below \SI{12}{\pico\second} at telecom wavelengths \cite{korzh_demonstrating_2018, zadeh_single-photon_2018, caloz_intrinsically-limited_2019}, and dark count rates below 0.5 counts per hour \cite{wollman_uv_2017, hochberg_detecting_2019}.  Emerging applications such as focal plane arrays for MIR astronomical spectroscopy \cite{the_ost_mission_concept_study_team_origins_2018}, detectors for dark matter search efforts \cite{hochberg_detecting_2019}, telecom LIDAR\cite{mccarthy_kilometer_2013}, and quantum imaging would benefit from SNSPDs' high efficiency and low intrinsic dark count rate performance for low photon-flux signals over wide wavelength ranges. However, these applications also require kilopixel to megapixel arrays covering millimeter-scale active areas. Traditional SNSPD readout uses high bandwidth RF cables for each channel of an array, but this places an unsustainable heat load on the cryogenic stage as the number of cables scales beyond a few tens of lines. Arrays of 64 pixels have been demonstrated using direct readout \cite{shaw_superconducting_2017} but scaling to the kilopixel range is challenging.  To overcome this limitation, various multiplexing schemes have been demonstrated which enable imaging capability while reducing the number of RF cables.  One approach uses a row-column biasing scheme \cite{allman_near-infrared_2015, wollman_kilopixel_2019, miyajima_single-flux-quantum_2019} and correlations between the detections on row and column readout channels to determine which pixel registered a detection. This architecture is able to read an \(N \times N\) array with \(2N\) readout lines. However, this architecture suffers from current redistribution among the pixels of each bias line, limiting the maximum count rate and degrading the timing jitter \cite{wollman_kilopixel_2019}. Multiplexing based on SFQ readout has been demonstrated in both standard  \cite{miyajima_high-time-resolved_2018} and row-column \cite{miyajima_single-flux-quantum_2019} arrays, but this approach still struggles with the heat load placed on the cryogenic stage.  A third approach fashions the active nanowire element as a delay line and uses differential readout to infer the location of absorption along the length of the nanowire \cite{zhao_single-photon_2017, zhu_scalable_2018}. This design requires the use of large impedance matching tapers to improve signal quality \cite{zhu_superconducting_2019}, but this adds significant kinetic inductance to the device which reduces the maximum count rate (MCR).  Furthermore, the spatial resolution is ultimately limited by the timing jitter of the system, necessitating the use of high performance amplifiers and readout electronics.  As an alternative, the microwave kinetic inductance detector (MKID) community has used frequency multiplexing to develop detector arrays with several to tens of kilopixels using few microwave feedlines at the cryogenics stage \cite{van_rantwijk_multiplexed_2016, meeker_darkness:_2018, szypryt_large_2017}.  While this type of frequency multiplexing has been demonstrated in SNSPD arrays using both DC \cite{sinclair_demonstration_2019} and AC \cite{doerner_frequency-multiplexed_2017} nanowire biasing, significant challenges remain in successfully scaling these architectures to the kilopixel scale.

We propose and demonstrate a new method of multiplexing, the thermally-coupled row-column (thermal row-column or TRC), which uses thermal coupling between two active SNSPD layers with channels arranged in rows and columns.  By measuring coincidence events between row and column channels, the location of absorption is inferred to be the intersection of those channels, forming a pixel.  The channels are electrically isolated at low frequencies, so there is no current redistribution and loss of electrical signal as occurs in the electrically coupled row-column architecture \cite{allman_near-infrared_2015, wollman_kilopixel_2019}.  The TRC has the additional advantage of not requiring  biasing resistors or wiring within the device active area, as needed for electrical row-column arrays, which increases the maximum active fill factor which can be achieved.

\begin{figure*} 
\includegraphics[width=\textwidth]{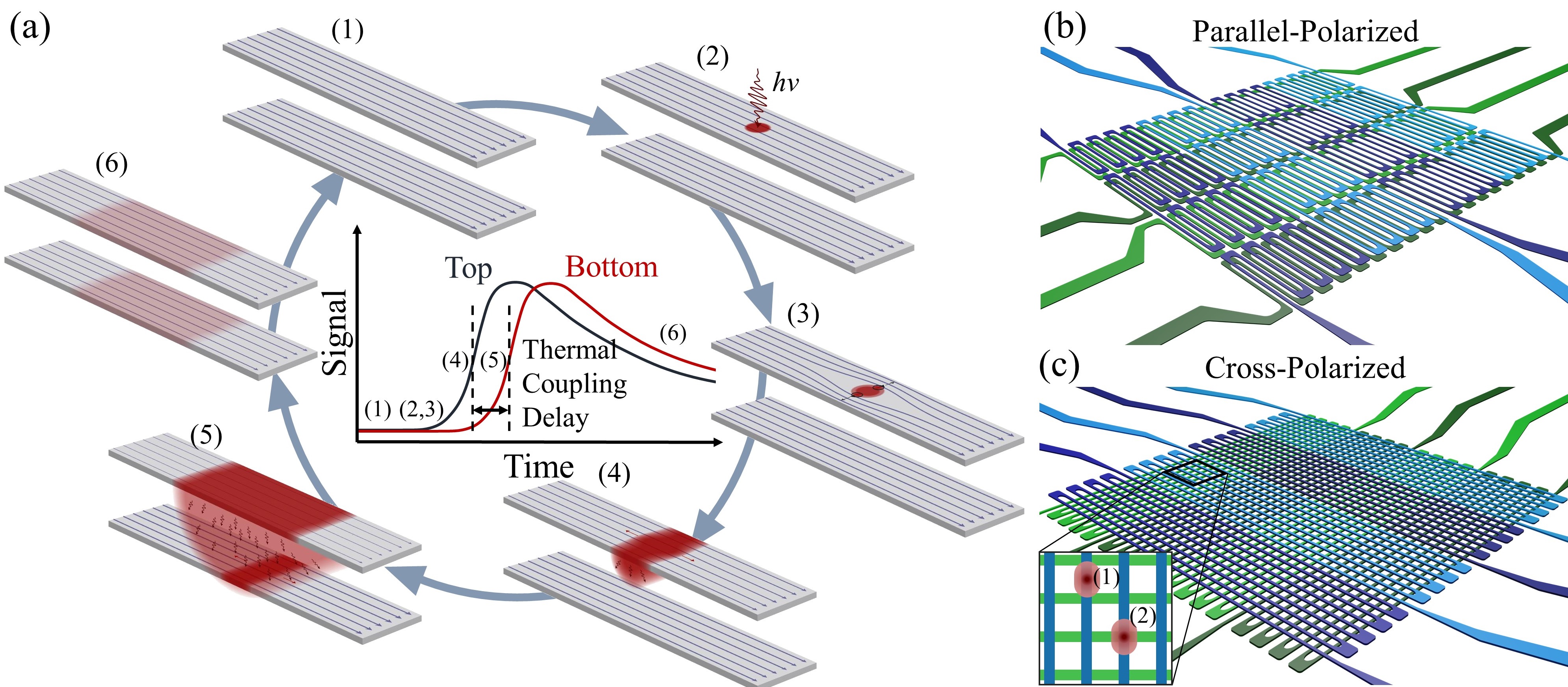}
\caption{(a) Detection process in TRC arrays.  The process begins with both nanowires current biased (1) until a photon is absorbed in one nanowire, creating a hotspot of excited quasiparticles (2).  As the hotspot evolves, a vortex-antivortex pair unbinds (or vortex entry from the edge for photons absorbed near the edge of the nanowire) (3), leading to the formation of a normal domain. As the normal domain grows along the length of the nanowire, (4) heat is coupled to the dielectric spacer, eventually increasing the temperature of the second nanowire and causing it to switch (5).  After current is diverted, both nanowires relax back to the superconducting state (6). Schematic illustration of \(4\times4\) (b) parallel-polarized and (c) cross-polarized thermal row-column devices.  The top layer channels (blue) are formed as columns while the bottom layer channels (green) are designed as rows.  The area where a column overlaps a given row forms a pixel, so the \(4\times4\) arrays shown here have 16 pixels.  The inset of (c) shows a top view of the overlap of the top and bottom nanowires for the cross-polarized design, highlighting how the distance heat must travel to trigger the second layer is greater for absorption location (1) compared to absorption location (2). As the nanowire fill factor increases, this variability decreases.}
\label{Fig: Device Schematics}
\end{figure*}

After the initial formation of a normal domain during photodetection, the inductance of the nanowire combined with Joule heating leads to dissipation of thermal energy.  Typically, this thermal energy is a nuisance to design and can limit the minimum allowable reset time of a detector before latching occurs \cite{kerman_electrothermal_2009}. In multi-element devices, this thermal energy can lead to crosstalk between adjacent pixels \cite{shaw_superconducting_2017} which limits the maximum achievable fill factor of the nanowires. The TRC utilizes this energy as a means of coupling row and column channels in an array. \(N\) nanowires are patterned into rows on one layer and \(N\) nanowires are patterned as columns on the second layer, with each nanowire having its own readout.  The heat generated during a photon detection on one nanowire layer raises the temperature of the dielectric and second nanowire layers, causing the nanowire on the second layer to switch, as shown in Figure \ref{Fig: Device Schematics}a. This effect was exploited in a single-pixel multilayer device where thermal coupling was used to trigger an avalanche between two nanowires in parallel, increasing signal output and reducing electrical noise jitter\cite{verma_athermal_2016}. While the dielectric spacer between the nanowires was only \SI{12}{\nano\meter} in that work \cite{verma_athermal_2016}, crosstalk in arrays \cite{shaw_superconducting_2017} suggests that thermal coupling should be observable over several hundred nanometer distances. By using the correlations between the detections on rows and columns, the location of the photon absorption is determined. Groups of channel detection events are considered correlated when they occur within a chosen time-span, known as the coincidence window. As with the standard row-column architecture, \(2N\) readout channels are required to operate an \(N \times N\) array. The thermally-coupled architecture is not limited to row-column style imaging arrays. As long as a unique overlap mapping exists between the channels of the layers, the same degree of multiplexing can be achieved, making the architecture viable for linear array spectrometers.

The nanowire orientation can be tailored to suit the needs of a particular application.  In the parallel-polarization design shown schematically in Figure \ref{Fig: Device Schematics}b, nanowires on both layers are oriented in the same direction and patterned directly above one another.  This is ideal for ensuring efficient and fast thermal coupling between the layers, but leads to increased capacitance between pixels which can increase electrical crosstalk. When embedded in a typical optical cavity, the parallel design maintains a strong polarization sensitivity.  For applications where this is undesirable, a cross-polarized design can be employed where the nanowires of the two layers are oriented perpendicularly as shown in Figure \ref{Fig: Device Schematics}c.  This can lead to polarization insensitive optical coupling, and the multilayer design enables broadband absorption with the appropriate optical stack design \cite{verma_three-dimensional_2012}. However, with the cross-polarized design, nanowires are not directly aligned, so the distance heat must travel to trigger the second layer depends on the absorption location.  This leads to additional jitter in the thermal coupling time, requiring a larger coincidence window which can impact the MCR (see Supporting Information). Despite this limitation, the overall timing jitter does not suffer in this configuration because the detection jitter is linked only to the first detector to `click.'

Two \(4 \times 4\) TRC arrays were fabricated; the first uses parallel polarized nanowires as shown schematically in Figure \ref{Fig: Device Schematics}b and the second uses perpendicularly polarized nanowires as in Figure \ref{Fig: Device Schematics}c.  Both devices use \SI{160}{\nano\meter} wide WSi nanowires on a \SI{1200}{\nano\meter} pitch covering a total active area of \SI{91.2}{\micro\meter} by \SI{91.2}{\micro\meter}. The full dimensions of the device stack and fabrication details are described in the Supporting Information. All measurements were performed using a temperature-controlled \SI{900}{\milli\kelvin} stage of a He-3 sorption fridge.  Each of the 8 channels of the arrays was individually biased and measured using a DC coupled cryogenic amplifier chain with an additional room temperature low noise amplifier.  Read-out was performed using 8 channels of a 64 channel time-to-digital converter (TDC) with a comparator front end \cite{shaw_superconducting_2017, wollman_kilopixel_2019}. Timetags were saved and analyzed in post-processing, but a real time streaming system could be developed using FPGA based readout.

\begin{figure*}
\centering
\includegraphics[width=\textwidth]{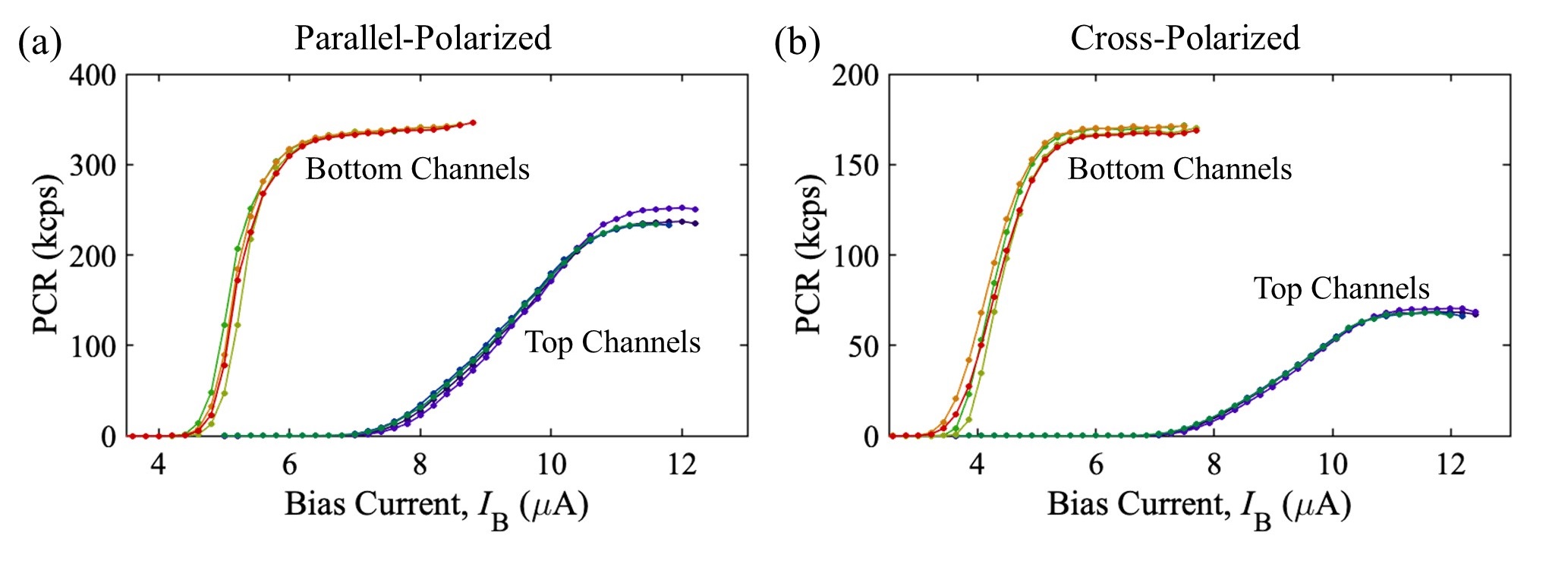}
\caption{Flood illumination photoresponse count rate (PCR) curves at \SI{1550}{\nano\meter} for (a) parallel-polarized and (b) cross-polarized devices when illuminated in the TM polarization with respect to the top nanowire layer.  The lower switching current of the bottom layer channels is attributed to inconsistencies in the fabrication of the two layers.  Measurement of the photoresponse was conducted with only the top or bottom channels biased at a time to avoid coupling between layers.  The small increase in count rate of the bottom layer of the parallel-polarized device near the switching current is attributed to increased photo-sensitivity of the meander bends, which occupy approximately 18\% of the active area.  Note that this is not present in the cross-polarized device where bends are only located at the edge of the active area.}
\label{Fig: PCR}
\end{figure*}

Both styles of device were characterized at a wavelength of \SI{1550}{\nano\meter} in a free-space coupled cryostat. The cryogenic windows of the cryostat include blackbody filters with a total transmission estimated to be greater than 96\% at \SI{1550}{\nano\meter} \cite{shaw_superconducting_2017}. The channels of the bottom layer exhibit a lower switching current than the top layer due to inconsistencies in the fabrication of the two nanowire layers.  Despite this, both layers show saturation of internal efficiency for all channels as shown in Figure \ref{Fig: PCR}. The number of counts is not distributed evenly between the top and bottom layers due to the properties of the optical cavity, but this distribution is reasonably consistent with rigorous coupled wave analysis (RCWA) modeling of the devices (see Supporting Information).

The TRC concept relies upon efficient thermal coupling between the active layers in order to extract the location where the detection occurred.  This requires that every detection on one layer triggers a detection on the other layer. We define the coupling efficiency \(p_i\left(I_\text{B,Top},I_\text{B,Bot}\right)\) as the probability that a `click' generated on a channel \(i\) leads to an unambiguous corresponding `click(s)' on the second layer. An ideal device would only generate a single corresponding detection on the second layer, making the readout as straightforward as possible.  If the additional complexity can be handled by the readout processing, triple coincidences (events with detections on three channels within a coincidence window, see Supporting Information) can be tolerated with the understanding that the detection occurred at the region where the pixels meet. This makes the implicit assumption that the probability of two photons being detected at the same time is negligibly small. The coupling probability for the two array designs was characterized using a \SI{1550}{\nano\meter} mode-locked laser for a range of bias currents on both top and bottom layers. For both array types, as the bias currents on both layers approach the switching current, near-unity coupling efficiency is achieved, demonstrating correct operation of the TRC architecture. 

\begin{figure*}
\centering
\includegraphics[width=\textwidth]{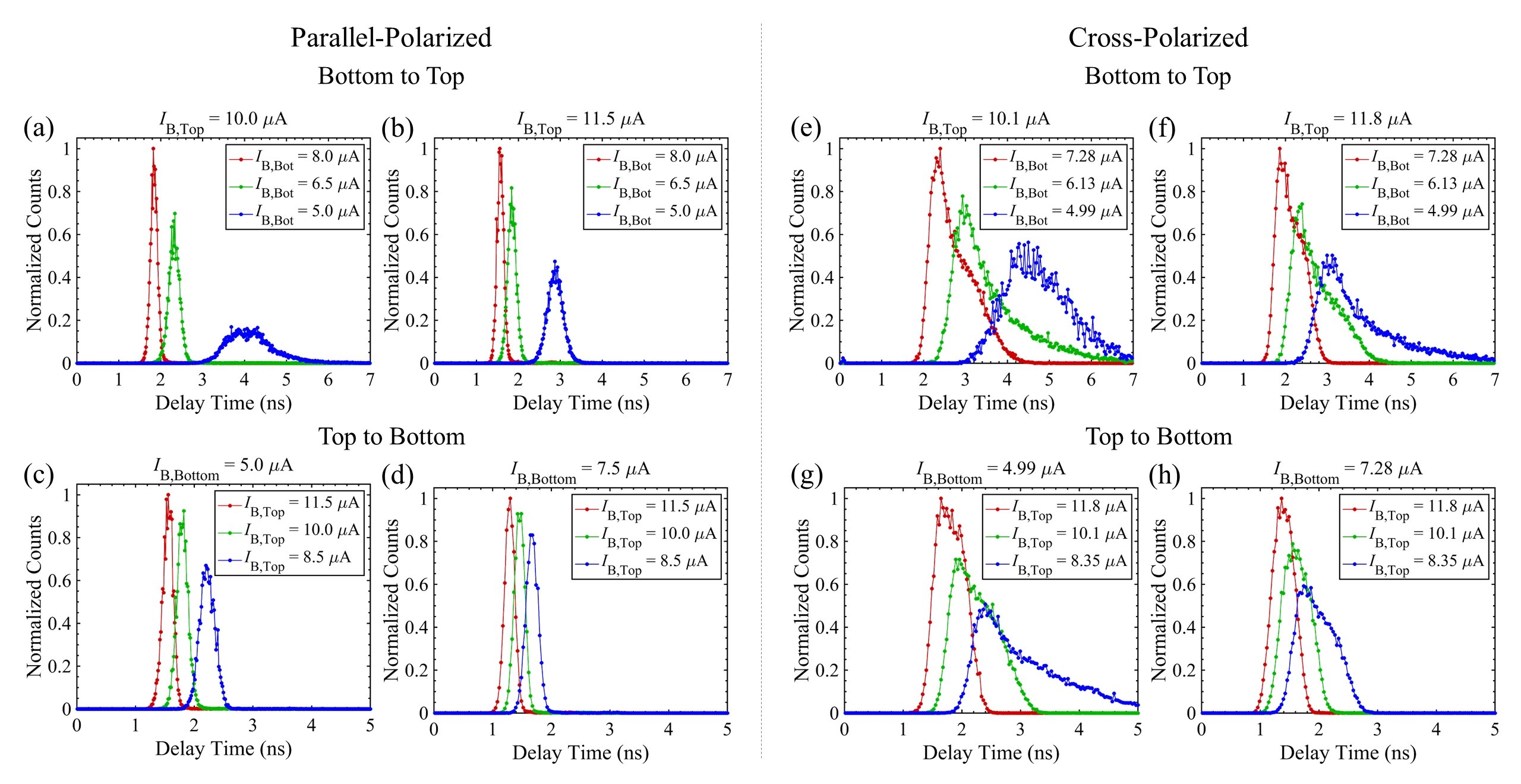}
\caption{Thermal coupling delay histograms for parallel-polarized (a-d) and cross-polarized devices (e-h).  The coupling efficiency increases as both the source and receiving channel bias currents increase.  Coupling from the bottom to top layer is less efficient than top to bottom due to the lower overall bias current and lower Joule heat generated during a detection.}
\label{Fig: Coupling Delay Histograms}
\end{figure*}

Thermal coupling between layers is not instantaneous.  There is a delay on the order of several nanoseconds between associated detection events due to the finite time required for heat to propagate from one nanowire to the second layer.  Furthermore, fluctuations in the heat transfer process lead to additional jitter in the detection timing of the second event compared to the first.  The coupling delay time is defined as the time difference between the initial detection event and the detection on the second layer.  Coupling delay time histograms are shown in Figure \ref{Fig: Coupling Delay Histograms} for a variety of bias conditions for the two arrays.  As can be expected based on the geometry, the parallel-polarized array demonstrates faster coupling and a narrower distribution of coupling times between layers than the cross-polarized device.  Differences in the distance heat must propagate to switch the second nanowire layer in the cross-polarized device lead to the appearance of a shoulder and extended long-delay time tail in the coupling behavior.  While this effect is diminished at higher bias currents, it is not eliminated completely. Consequently, when compared to the parallel-polarized device, the cross-polarized design requires a wider coincidence window when defining correlated detection events.  See the Supporting Information for additional details regarding thermal coupling.

\begin{figure*} 
\centering
\includegraphics[width=\textwidth]{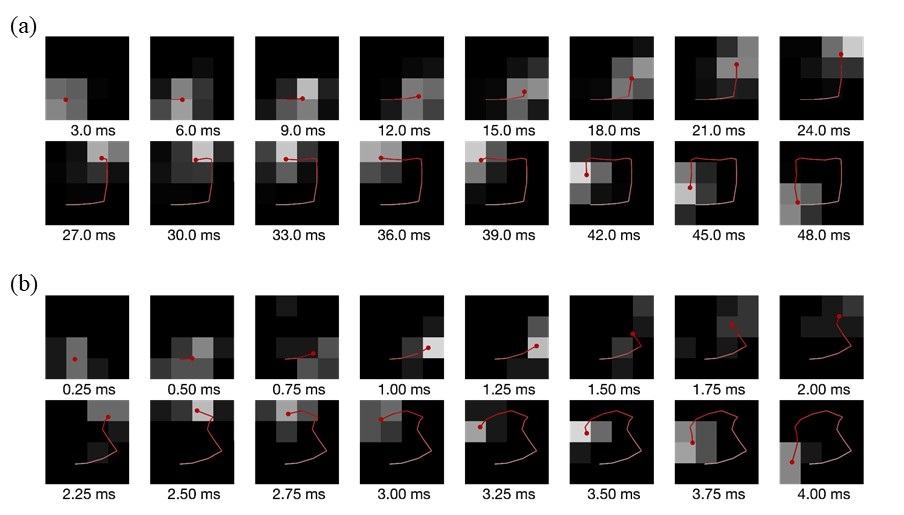}
\caption{Laser spot tracking in a square pattern using the cross-polarized array.  (a) Tracking using a \SI{48}{\milli\second} period and \SI{3}{\milli\second} frames with an average of 146.5 photon detections per frame. (b) Tracking using a \SI{4}{\milli\second} period and \SI{250}{\micro\second} frames with an average of 11.7 photon detections per frame. The red circle indicates the centroid of the detections for a given frame and the line shows the path of the centroid for the sequence of frames.}
\label{Fig: Spot Tracking}
\end{figure*}

To demonstrate the imaging capability of the TRC devices, we translated a laser spot on the active area of the device using a fast steering mirror. For the cross-polarized device, the spot was directed in a square pattern on the device with varying periods, and the location of the spot was extracted using the centroid of the measured counts.  The spot tracking was reliably demonstrated up to limits of the steering mirror speed with period of \SI{4}{\milli\second} using \SI{250}{\micro\second} frames, with an average of 11.7 photon detections per frame.  Figure \ref{Fig: Spot Tracking} shows representative sample frames demonstrating this tracking capability, but based on the MCR performance of the device (see Supporting Information), a similar level of tracking accuracy could be achieved at a \SI{100}{\micro\second} period with a photon flux increase by a factor 50 with minimal blocking loss.

The TRC architecture is not limited to the WSi material system. For devices which require lower jitter and higher count rates, NbN or NbTiN arrays are possible.  The challenge with using these higher \(T_c\) materials is ensuring sufficient thermal coupling between the two layers to generate thermally-induced switching. The greater \(T_c\) requires that the temperature of the second nanowire layer must be elevated to a higher temperature to switch the device when biased at a given fraction of the nanowire depairing current.  While this is not expected to be a problem for the parallel-polarized design, cross-polarized devices may require a high fill factor in order to ensure that there is sufficient overlap between the layers for efficient coupling. A better ability to model heat transfer in thin dielectrics and superconductors is needed to reliably predict the thermal coupling between superconducting nanowire layers in TRC devices. 

In conclusion, we have demonstrated a row-column style multiplexing architecture for SNSPD imaging arrays using thermal coupling between two active layers.  This new architecture eliminates the current-redistribution and limited fill factor shortcomings of electrically coupled row-column SNSPD arrays. Similar device structures using this platform provide a new means of experimentally probing low temperature heat transfer properties of thin dielectric structures and can be used to produce optimized thermal row-column arrays in the future. The 16-pixel devices characterized in this work are immediately scalable to the kilopixel size for use with existing readout systems and, when integrated with optical cavities, are expected to achieve optical efficiency exceeding 85\%. When combined with an additional multiplexing scheme through time domain or standard electrical row-column mechanisms, it will be possible to reach megapixel sized arrays with a manageable number of readout cables.  The thermal row-column promises to be a key enabling technology for achieving large area, low intrinsic dark count rate, and high efficiency photon counting imaging systems for low photon flux signals.

\section{Associated Content}
The Supporting Information includes the fabrication and design details, timing jitter characterization, details of the thermal coupling efficiency, thermal coupling delay times, maximum count rate characterization, and optical efficiency characterization.


\section{Acknowledgments}
The authors thank V. Verma, A. McCaughan, and S. W. Nam for helpful discussions. This research was performed at the Jet Propulsion Laboratory, California Institute of Technology, under contract with the National Aeronautics and Space Administration. This work was supported by a NASA Space Technology Research Fellowship.  Support was provided in part by the DARPA Defense Sciences Office, through the DETECT program.

\newpage 
\bibliography{SNSPD_References}

\end{document}


\copyright{2020. California Institute of Technology. Government sponsorship acknowledged.}

\section{Device Fabrication}
Devices were fabricated on a 4 inch silicon wafer.  A \SI{60}{\nano\meter} thick Au back reflector was patterned via optical lithography and lift-off before a \SI{155}{\nano\meter} thick film of SiO\(_2\) was sputtered to form an insulating layer.  An \SI{8}{\nano\meter} thick WSi film was sputtered from a compound target, after which Au contact pads and leads were patterned via lift-off. The bottom nanowire layer was pattered using electron-beam lithography and etched using an ICP RIE dry-etch of CHF\(_3\) and O\(_2\).  After the first nanowire layer was patterned, a buffer layer of \(\sim\)\SI{190}{\nano\meter} SiO\(_2\) was sputtered and smoothed using angled incidence Ar ion milling.  The sputtering and ion milling process was necessary to planarize the surface sufficiently to yield the second layer of nanowires without constrictions.  The second nanowire layer and leads were patterned using the same process as the lower layer and a final capping layer of \(\sim\) \SI{65}{\nano\meter} of SiO\(_2\) was sputtered for passivation.  Optical efficiency could be enhanced by designing and depositing antireflection coatings to form an optical stack.

\begin{figure*} 
\centering
\includegraphics[width=\textwidth]{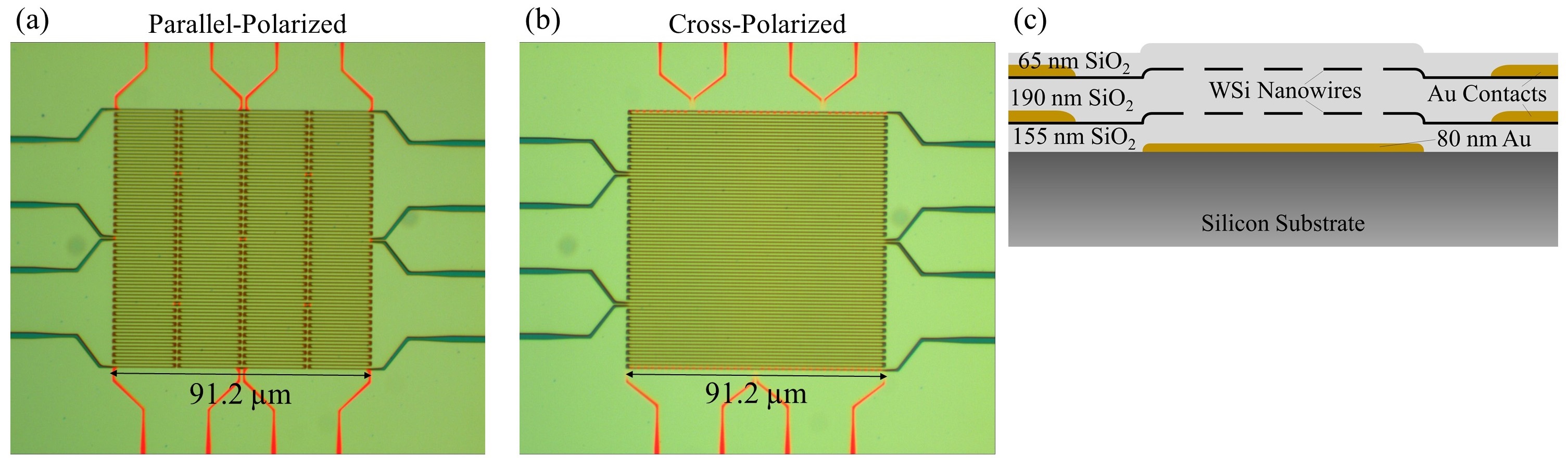}
\caption{Optical microscope images of (a) parallel-polarized and (b) cross-polarized devices. Blue channels are the top layer while red channels are the bottom layer.  (c) Schematic cross-section of the TRC devices showing layer thicknesses.}
\label{Fig: Device Design}
\end{figure*}

Two designs were fabricated using this procedure.  The first was a \(4\times4\) array using parallel nanowires on the top and bottom layers as shown schematically in Main Text Figure 1b and in the optical micrograph of Figure \ref{Fig: Device Design}a.  The total active area was \SI{91.2}{\micro\meter} by \SI{91.2}{\micro\meter} with each row and column consisting of four \SI{22.8}{\micro\meter} by \SI{22.8}{\micro\meter} units connected in series. The second design was a \(4\times4\) cross-polarized array using perpendicular top and bottom nanowires as shown schematically in Main Text Figure 1c and in the microscope image of  Figure \ref{Fig: Device Design}b.  The total active area was \SI{91.2}{\micro\meter} by \SI{91.2}{\micro\meter}.  Both devices use \SI{160}{\nano\meter} wide nanowires with \SI{1200}{\nano\meter} pitch.  The low fill factor was selected to avoid crosstalk between adjacent pixels \cite{shaw_superconducting_2017} while maintaining a uniform nanowire fill throughout the entire active area.  A higher nanowire fill-factor could be achieved while avoiding crosstalk by increasing the fill-factor within a pixel but leaving additional guard space between adjacent pixels.

Design of TRC arrays requires a balance between the electrical and thermal crosstalk mechanisms in neighboring nanowires.  During a detection event in these two-layer devices, both mechanisms couple energy from the detecting nanowire to adjacent channels on both the original and second layers\cite{verma_athermal_2016}.  Capacitive coupling is the dominant mechanism of microwave coupling in this geometry, so as the thickness of the spacer layer between the nanowire layers decreases, the strength of the electrical coupling between neighboring nanowires increases.  Electrical crosstalk of this form is undesirable because it is distributed across all channels of the second layer rather than localized to the detecting pixel. The channel to channel capacitance is estimated to be approximately \SI{31}{\femto\farad} for the parallel-polarized and \SI{24}{\femto\farad} for the cross-polarized devices fabricated in this work.  Thermal coupling is inherently local to where the detection occurs, but the timescale of this coupling is slow compared to electrical crosstalk.  The time difference between the generation of normal domains in the two detecting nanowires, defined as the thermal coupling delay time (see Main Text Figure 1a), is on the order of a few nanoseconds for the geometries we demonstrate in this work. 

\subsection{Timing Jitter}
\begin{figure*} 
\centering
\includegraphics[width=\textwidth]{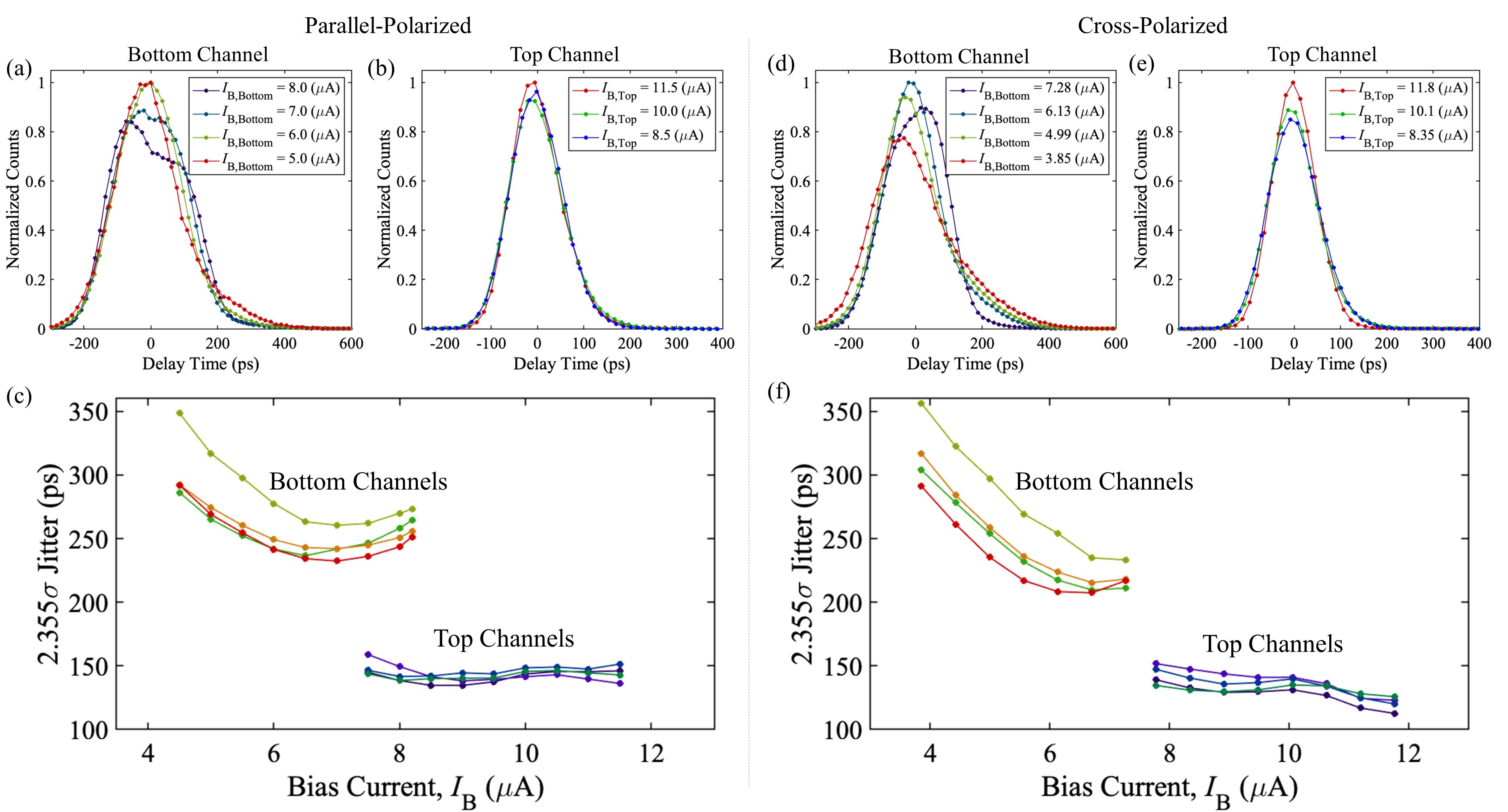}
\caption{Jitter characterization for channels of the (a-c) parallel-polarized and (d-f) cross-polarized arrays.  Jitter histograms for a representative channel of the bottom layer (a, d) show non-Gaussian IRF with a distinct asymmetry in the distribution at high bias currents.  We attribute this to electrical coupling between the nanowire layers and to the back reflector which reduces the signal propagation velocity and leads longitudinal geometric jitter\cite{zhao_single-photon_2017}. In contrast, the IRF of the top channels (b, e) is nearly Gaussian.  To accommodate the non-Gaussian behavior of the lower layer channels, the total jitter is defined by \(2.355 \sigma\) (c, f), and is significantly larger for the bottom channels than the top, even when biased at the similar currents where the electrical noise contribution to the jitter from the readout electronics is similar.  A device optimized for improved timing performance must manage the electrical coupling and signal propagation of the bottom layer to take full advantage of the low timing jitter of SNSPDs.}
\label{Fig: Jitter}
\end{figure*}

The timing jitter of the system was characterized using a \SI{20}{\mega\hertz} repetition rate \SI{1550}{\nano\meter} mode-locked laser.  A phase-locked loop circuit converted the electrical sync of the laser to a \SI{10}{\mega\hertz} clock which acted as the timing source for the time to digital converter (TDC). The laser was focused to a Gaussian spot with diameter of approximately \SI{85}{\micro\meter} FWHM.  Due to non-Gaussian instrument response function (IRF) of the array channels, the timing jitter is defined as 2.355\(\sigma\) where \(\sigma\) is the sample standard deviation of the photocount timetags modulo \SI{50}{\nano\second}.  Figure \ref{Fig: Jitter} shows the IRF and the bias current dependent timing jitter for representative channels of both layers of the array, demonstrating that the overall jitter is less than \SI{300}{\pico\second} at optimal bias regardless of which nanowire absorbs the photon. The jitter of a given pixel of the array is the combination of the two histograms of the channels forming the pixel weighted by the relative absorption efficiency of the two channels because the measured arrival time of the photon is given by the detection time of the first channel to detect the photon.

\section{Thermal Coupling Efficiency}
\begin{figure*}
\centering
\includegraphics[width=\textwidth]{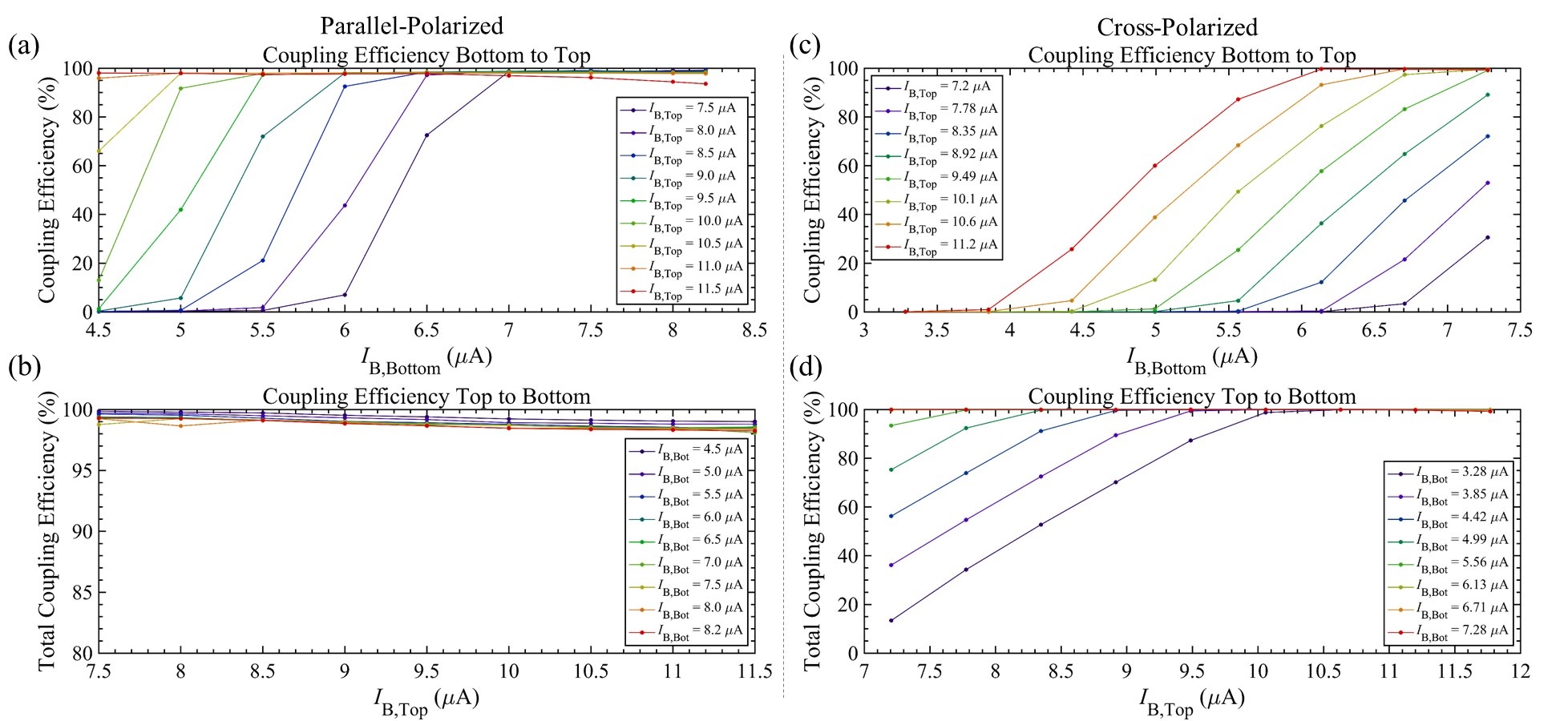}
\caption{Coupling efficiency for representative channels of the parallel-polarized (a, b) and cross-polarized arrays (c, d).  The coupling efficiency increases as both the source and receiving channel bias currents increase.  Coupling from the bottom to top layer is less efficient than top to bottom due to the lower overall bias current and lower Joule heat generated during a detection.  The transition from no coupling to efficient coupling is more gradual in the cross-polarized device compared to the parallel-polarized device due to the non-uniform overlap between the nanowires of the two layers.  Lines are to guide the eye.}
\label{Fig: Coupling Efficiency}
\end{figure*}

The coupling probability for the two array designs was characterized using a \SI{1550}{\nano\meter} mode-locked laser for a range of bias currents on both top and bottom layers.  All four channels of a given layer were biased at the same current and the photon flux was kept sufficiently low to make the probability of two photons being detected in the same optical pulse negligible.  For each channel, a bias dependent calibration delay was applied to all timetags such that the mean delay for photons detected on each channel is zero based on the timebase of the mode locked laser. The coupling probability is shown as a function of bias current in Figure \ref{Fig: Coupling Efficiency} for representative channels on both the bottom and top layers.  The parallel-polarization device exhibits near unity coupling efficiency for a wide range of bias current combinations \(\left(I_\text{B,Top},I_\text{B,Bot}\right)\) as is expected based on the direct overlap between the nanowire regions of the two layers (see Main Text Figure 1b).  The coupling efficiency increases to near unity as both the detecting and receiving nanowire bias currents increase.  The explanation is straightforward.  As the detecting nanowire bias current increases, more Joule heat is released during detection which heats the dielectric layer and second nanowire layer, increasing the coupling efficiency.  As the bias current on the receiving layer increases, a smaller change in nanowire temperature is required to trigger a switching event, making the nanowire more sensitive to small heat pulses generated by the source nanowire.

The cross-polarized device displays more varied behavior due to the nonuniform overlap between the two layers (see Main Text Figure 1c).  Compared to the parallel-polarization device, higher bias currents are required to ensure near unity \(p_{i}\left(I_\text{B,Top},I_\text{B,Bot}\right)\), but when both layers are biased near their switching currents, unity detection can be achieved. Furthermore, the transition region between no coupling and efficient coupling occurs over a wider range of bias currents.  This is consistent with the varying overlap between the two nanowire layers.

\begin{figure*}
\centering
\includegraphics[width=\textwidth]{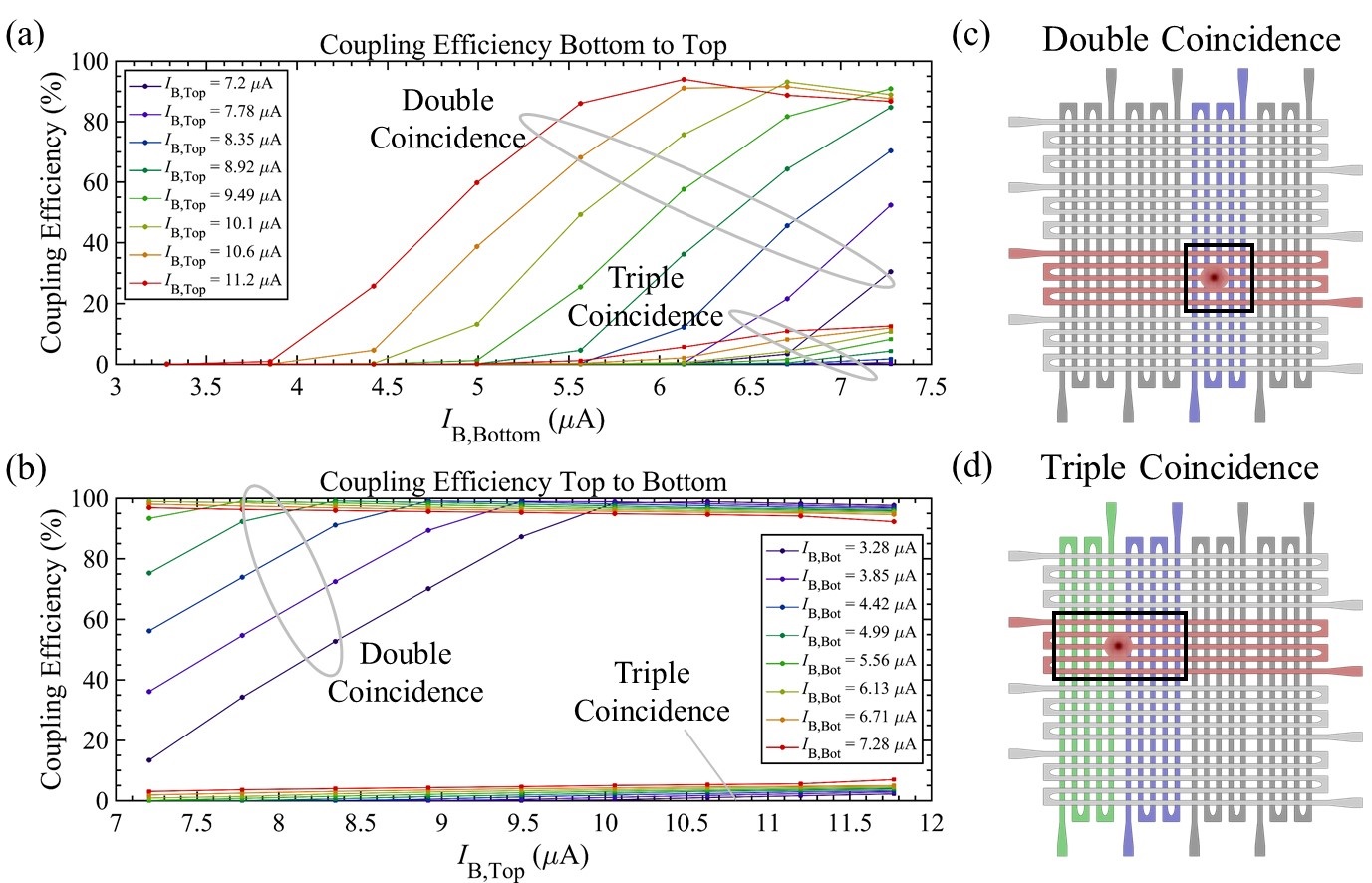}
\caption{Cross-polarized array coupling efficiency by coincidence number for (a) a bottom channel triggering a top channel and (b) a top channel triggering a bottom channel. The number of triple coincidences is non-negligible for high bias currents on the either the top or bottom channels.  Unambiguous double coincidence and single photon triple coincidence detection events are shown schematically in (c) and (d) respectively. The colored channels indicate those which register a detection during the photon detection event, and the overlap of the detecting channels indicates the determined location.}
\label{Fig: Coupling Numbers}
\end{figure*}

The total coupling efficiency between layers in the cross-polarized device increases monotonically as the bias current increases on either the detecting or receiving channels, but the fraction of double coincidences (two detection events within a coincidence window) and triple coincidences (three detection events) is not constant.  As seen in Figure \ref{Fig: Coupling Numbers}a, as the bias current on the bottom channel increases, the coupling efficiency for double coincidences, shown schematically in Figure \ref{Fig: Coupling Numbers}c, initially increases due to the increased Joule heat generated.  However, this coupling efficiency actually decreases at the highest bias currents of the top channel.  This occurs because the large amount of heat generated by the combination of the two channels switching can be sufficient to switch a neighboring channel which is part of a different pixel. The number of these triple coincidences, shown schematically in Figure \ref{Fig: Coupling Numbers}d, increases as both the top and bottom bias currents increase, but the bias current of the top channel plays a more significant role than that of the bottom channel due to the larger overall bias currents involved.  Triple coincidences are only considered to be valid if the the two detecting channels on the same layer are adjacent and the probability of two photons being absorbed within a coincidence window is negligibly small.  Analysis of the correlations confirms that at high bias current combinations, \(> 98\%\) of the triple coincidence events occur between two adjacent channels on one layer and a single channel on the second layer.

Triple coincidences are not inherently problematic in determining the location of the photon detection because at high bias current operating points, \(>\)99.5\% of triple coincidences occur with the first receiving channel belonging to the second layer of the device.  Thus, a proper detection location can be assigned based on the detecting and first receiving channels alone. However, to simplify an FPGA based readout system, it is desirable to have only double coincidence events.  Furthermore, the possibility of single photon triple coincidence detection events makes it impossible to distinguish between blocking caused by two photon detections and a single photon triple coincidence event. In a cross-polarized device, single photon triple coincidences can be minimized by leaving guard space between adjacent pixels.  In the parallel-polarized array, single photon triple coincidences make up a negligible fraction of the detection events for the large pitch arrays tested in this work.  

\section{Thermal Coupling Delay Time}
\begin{figure*}
\centering
\includegraphics[width=\textwidth]{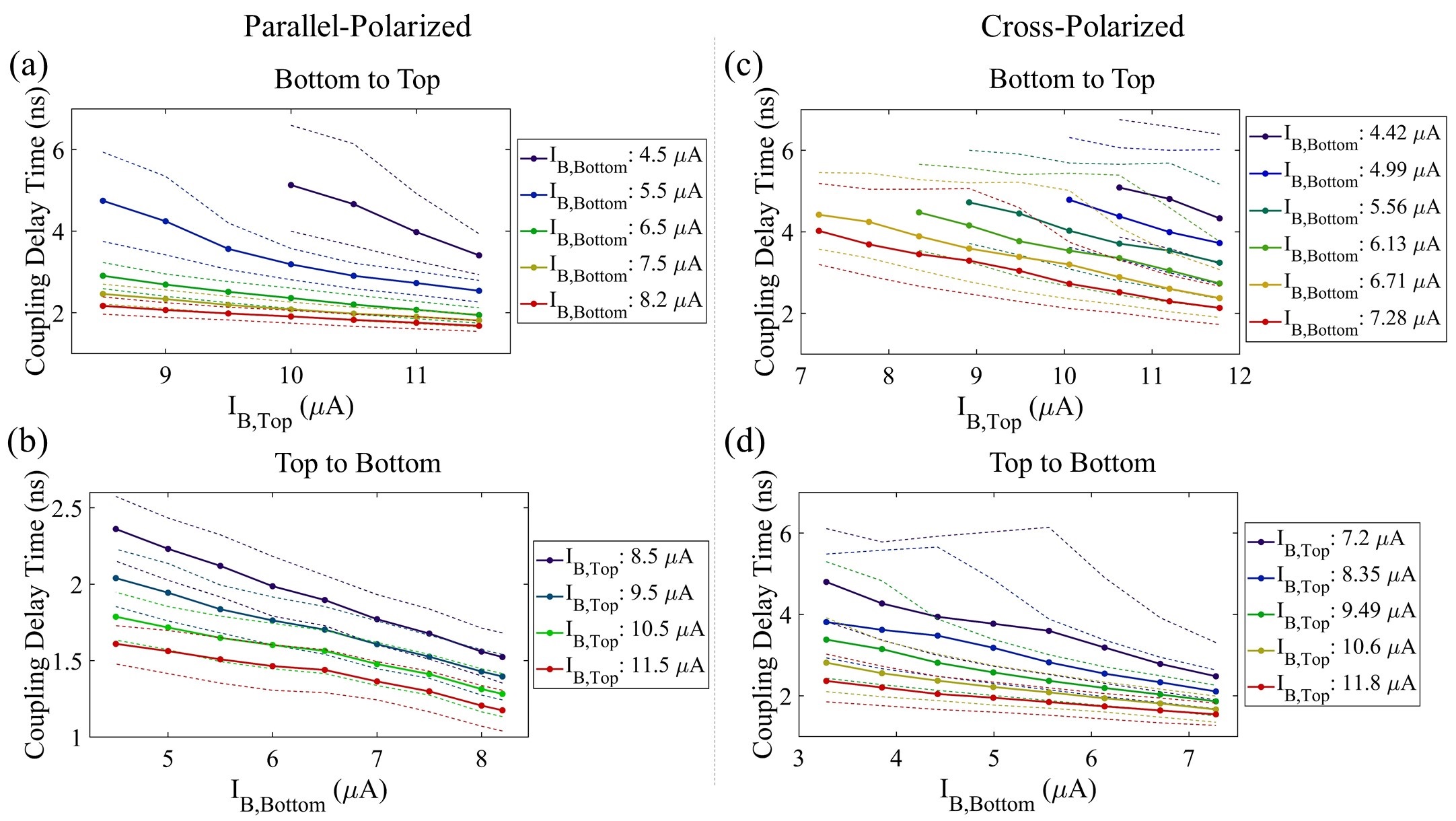}
\caption{Thermal coupling delay times for (a, b) parallel-polarized and (c, d) cross-polarized representative channels of the arrays.  Solid lines with symbols indicate the mean delay time while dashed lines indicate the 90\% distribution bounds to show the width of the coupling delay time distribution. The coupling times and the width of the coupling distribution both decrease as the bias current increases on either the source or receiving channels.  The cross-polarized channels show a wider distribution and longer mean coupling time than the parallel-polarized channels for comparable bias conditions.}
\label{Fig: Coupling Time}
\end{figure*}

The thermal coupling delay time is a function of the bias currents on both the detecting and receiving channels.  The average coupling delay times for various bias current configurations are shown in Figure \ref{Fig: Coupling Time} for both array designs.  Due to the direct overlap between layers, the parallel-polarized device demonstrates shorter coupling delay times and narrower distributions compared to the cross-polarized device for comparable values of the bias currents. The longer delay times of the cross-polarized device require the use of a longer coincidence window when processing timetags to determine row and column coincidences.  

As the bias currents increase, the delay time decreases and the distribution width becomes narrower. Despite separation distances of only a few hundred nanometers between layers, the delay times are on the order of several nanoseconds.  This is significantly slower than expected based on ballistic propagation of phonons and slower than expected based on thermal modeling using a Casimir limited thermal conductivity for thin film dielectrics.  There is experimental evidence that this type of suppressed thermal conductivity in thin film SiO\(_2\) is typical \cite{cahill_thermal_1989, allmaras_thin-film_2018}, but more advanced modeling is needed to fully characterize the behavior, then use the results to optimize the thermal design of TRC arrays. Specialized devices can be designed which use a variety of nanowire widths, inductances, and overlap spacings in order to build a detailed experimental dataset of the coupling delay times between nanowires under different geometries.  This type of device offers a new means of studying this physics and improving both engineering and fundamental models of this process.  Previous work demonstrated that a complete electrothermal model of the reset dynamics of WSi nanowires requires an accurate model of the heat transfer in the surrounding dielectric material \cite{allmaras_thin-film_2018}, but the TRC architecture offers a way of experimentally studying this process in order to develop and validate such a model. This understanding is necessary to fully optimize TRC arrays.
 
\section{Maximum Count Rate}
The maximum count rate (MCR) was characterized using a \SI{1550}{\nano\meter} CW source with variable attenuators and a focused spot with a FWHM diameter of \SI{85}{\nano\meter} on the array.  The MCR is defined as the count rate where the efficiency of the detector drops by 3 dB compared to the low count rate efficiency.  The MCR curves are shown below in Figure \ref{Fig: MCR} for both array designs.  The parallel-polarized array demonstrates an MCR of \SI{14.6}{\mega cps} while it is only \SI{10.8}{\mega cps} for the cross-polarized array.  The parallel-polarized array was biased at \SI{11.5}{\micro\ampere} and \SI{8.2}{\micro\ampere} for the top and bottom channels respectively while the cross-polarized array was biased at \SI{11.8}{\micro\ampere} and \SI{7.28}{\micro\ampere} for the top and bottom channels respectively.  Due to the polarization dependent relative efficiency of the channel layers, the maximum count rate of the cross-polarized device is slightly polarization dependent. When photons are preferentially absorbed in either the top or bottom layer, the array experiences elevated blocking loss, resulting in a reduced MCR. The MCR depends on the coincidence window chosen to analyze the timetag data.  For the parallel-polarized data, the optimal window is \SI{3.5}{\nano\second} while for the cross-polarized array, the optimal window is \SI{4.5}{\nano\second}.  The larger optimal window for the cross-polarized array is consistent with the longer thermal coupling delay time for these bias currents.  In the MCR analysis of the cross-polarized array, triple coincidences were not included as unambiguous counts because the photon flux is not sufficiently low to eliminate the possibility of multiple photons being detected within a coincidence window.  Single-photon triple coincidences lead to additional blocking loss and are the primary cause of the lower MCR in the cross-polarized array compared to the parallel-polarized device.

\begin{figure*}
\centering
\includegraphics[width=\textwidth]{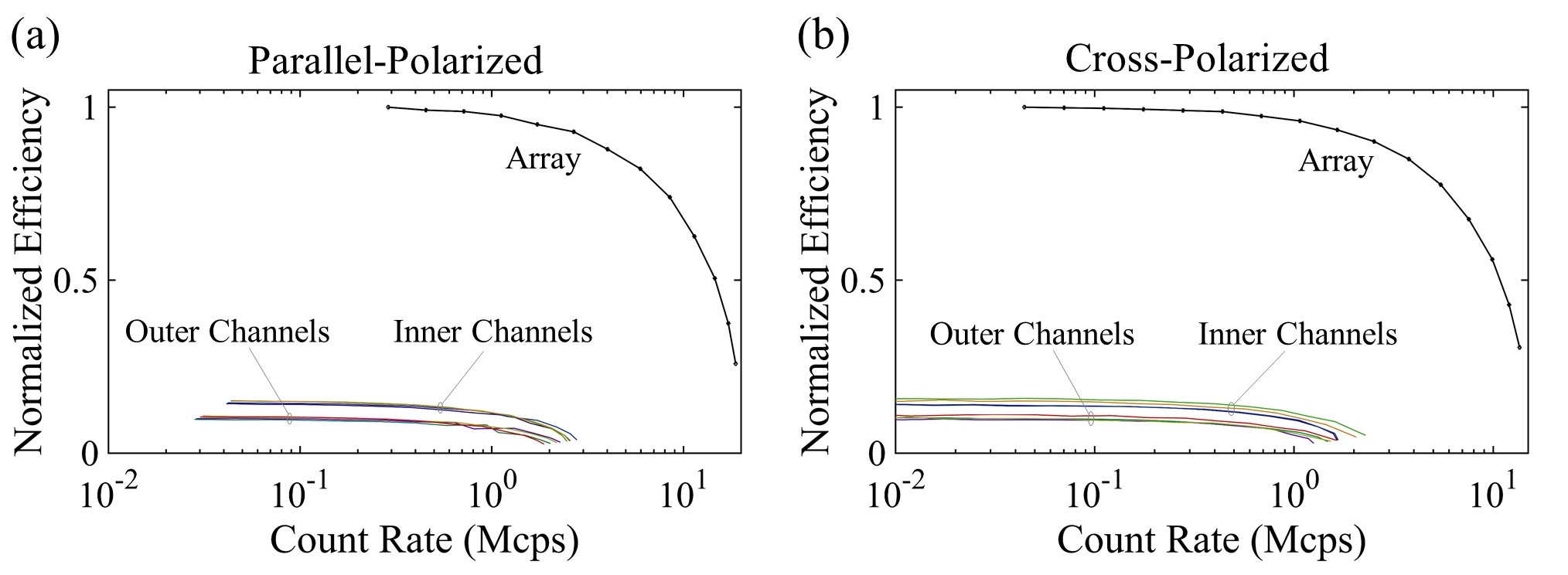}
\caption{MCR curves for (a) parallel-polarized and (b) cross-polarized arrays.  The parallel-polarized device was illuminated with TE polarized light while the cross-polarized array was illuminated with linear polarization rotated \SI{45}{\degree} with respect to the TE and TM modes.  For both arrays, the two inner channels of each layer show higher count rates than the two outer channels due to unequal illumination with the laser spot.  In these figures, the parallel-polarized device uses a coincidence window of \SI{3.5}{\nano\second} while the cross-polarized array uses a coincidence window of \SI{4.5}{\nano\second}.}
\label{Fig: MCR}
\end{figure*}
 
The thermal coupling delay times can have a direct impact on the MCR which can be achieved over the array without position ambiguity. Two factors contribute to position ambiguity. The first is due to blocking loss. During the deadtime of a channel, detections which occur on the other layer but overlapping the dead channel will not have a correlated click.  In the limits of equal illumination of pixels, ideal thermal coupling, identical channel properties, and Poisson distributed photons, the array MCR (\(MCR_A\)) scales according to \(MCR_A \sim N^2\:MCR_i/(2N - 1) \) due to this deadtime, where \(MCR_i\) is the channel MCR. The second factor is associated with the timing uncertainty of correlated detections.  Electrical and thermal timing jitter lead to a range of thermal coupling times between the photon-induced detection and the thermally-coupled detection.  When analyzing the channel detection times to determine the coincidence events, one must define a range of times during which two events can be correlated, known as a coincidence window. Two detection pairs which occur within the same coincidence window lead to ambiguity as to which combination of pixels were the source of detection events because two rows and two columns have four potential pixel locations. As the array size becomes large, the coincidence window of correlated detections limits the counting rate. A large thermal coupling delay indicates that the rate of temperature increase in the second nanowire is slow.  Therefore, fluctuations during thermal coupling process lead to larger fluctuations in the coupling delay time, analogous to the dependence of electrical noise jitter on signal slew rate. This requires setting a larger coincidence window, which is only acceptable for low count-rate applications.  Reducing the thickness of the spacer layer quickens thermal coupling and potentially leads to a smaller coincidence window, but leads to additional electrical coupling.  An optimized design would reduce the thickness of the dielectric spacer to the minimum thickness where the electrical crosstalk can still be tolerated. 
 
\section{Optical Efficiency}
The optical efficiency of each device was measured using a \SI{1550}{\nano\meter} CW laser focused to a Gaussian spot with a FWHM diameter of approximately \SI{33}{\micro\meter}. Only counts which exhibit unambiguous coincidence groups are included in the efficiency measurement.  This includes triple coincidences for the cross-polarized device as defined above. We find that the rigorous coupled wave analysis (RCWA) technique provides a reasonable prediction of device efficiency. The parallel polarization device exhibits 34\% TE and 11\% TM efficiency while the RCWA calculation predicts 34\% TE and 14\% TM efficiency. In the TM polarization, the model predicts a 64:36 bottom to top layer absorption ratio which is close to the experimentally measured 58:42 ratio. The cross-polarized device has 30\% TE and 26\% TM efficiency while the RCWA calculation predicts 30\% TE and 29\% TM efficiency for polarization with respect to the top layer orientation. In the TM polarization, the model predicts a 69:31 bottom to top ratio compared to the measured 72:28 ratio. The parallel-polarized device incurs additional optical coupling losses compared to the cross-polarized design due the presence of sections of non-photosensitive meander bends comprising approximately 18\% of the active area.  This advantage of the the cross-polarized architecture is significant for small pixel sizes, but becomes less important as the pixel size increases or the illuminating spot is not focused on an area of the array where the bends are present. 
 
\newpage 
\bibliography{SNSPD_References}